\documentclass{PoS}
\usepackage{wrapfig}

\title{Gamma-rays from SNIa}

\ShortTitle{Gamma-rays from SNIa}

\author{\speaker{J. Isern}\\
        Institut de Ci\`encies de l'Espai (CSIC) \& Institut d'Estudis Espacials de Catalunya (IEEC)\\
        E-mail: \email{isern@ice.cat}}

\author{E. Bravo\\
        ETSAV, Universitat Polit\`ecnica de Catalunya\\
        E-mail: \email{eduardo.bravo@upc.edu}}

\author{P. Jean \\
       IRAP \& Universit\'e de Toulouse\\
        E-mail: \email{pjean@irap.omp.eu}}

\abstract{Type Ia supernovae are thought to be the outcome of the thermonuclear explosion of a carbon/oxygen white dwarf in a close binary system. Their optical light curve is powered by thermalized gamma-rays produced by the radioactive decay of 56Ni, the most abundant isotope present in the debris. The maximum and the shape of the light curve strongly depends on the total amount and distribution of this freshly synthesized isotope, as well as on the velocity and density distribution of the ejecta. Gamma-rays escaping the ejecta have the advantage of their lower interaction with the ejecta, the possibility to distinguish among isotopes and the relative simplicity of their transport modelling, and can be used as a diagnostic tool for studying the structure of the exploding star and the characteristics of the explosion, as it has been proved in the case of SN2014J.}

\FullConference{Frontier Research in Astrophysics II\\
		23-28 May 2016\\
		Mondello (Palermo), Italy}

\begin{document}

\section{Introduction}
Type Ia supernovae (SNIa) are the outcome of the 
thermonuclear explosion of an accreting carbon--oxygen white dwarf in a binary 
system. Traditionally two possibilities have been considered, the central ignition
 of the white dwarf when it approaches to the Chandrasekhar's mass, or the
explosion of a C/O white dwarf of arbitrary mass triggered by the detonation 
of a freshly accreted helium layer.  Although the first scenario \cite{bran81}
 has been the preferred one for many years, the situation is not clear at 
present. The recent discovery of new SNIa subtypes has 
forced to abandon the idea that a unique explanation is possible. As a 
consequence, several new scenarios and explosion mechanisms have been 
advanced (prompt and delayed coalescence of double degenerates, frontal 
collisions of degenerates in clusters...) besides the classical ones, as well 
as different mechanisms of ignition and propagation of the flame leading to 
structures that in some cases are far from the simple spherical symmetry 
considered up to now.

The maximum brightness and shape of the optical SNIa light curves are 
determined by the distribution of the freshly synthesized $^{56}$Ni (and $^{56}$Co), as well as by the velocity and density profiles of the ejecta. 
The problem is that the physics of the radiative transfer in such expanding 
envelopes is so complex that the reproduction of the spectrum does not 
guarantee the correct description of the light curve. Therefore, since the 
most direct outcomes of the process are {\it the freshly synthesized 
radioactive ashes}, $\gamma$--ray astronomy can provide the deepest observational 
insight to the problem of SNIa \cite{clay69,gome98,the14}.

\section{The expected $\gamma$--ray emission}

The thermalization of the gamma--photons produced by the
radioactive decay of $^{56}$Ni and $^{56}$Co is the engine that powers the 
light curve of SNIa. As ejecta expands, more and more photons avoid 
thermalization and escape and can be used as a diagnostic tool. The expected 
$\gamma$--ray energy ranges from few keVs to 4 MeV approximately,
The different scenarios and burning modes lead to differences in the intrinsic
properties of the ejecta like the density and velocity profiles or the
amount and distribution of the radioactive material synthesized. This, in turn
translates into differences in the line width and light curve of the expected
 $\gamma$--ray emission. Thus,
the observation of a type Ia supernova in the $\gamma$--ray light 
becomes a privileged diagnostic tool respect to other measurements. This is so
 because the penetrating power of high energy photons and the association of 
$\gamma$--ray lines
with specific isotopes can provide information about deep layers of the ejecta
 even at early epochs as well as powerful constraints on the nucleosynthesis.

Several scenarios have been advanced up to now: i) In the single degenerate scenario (SD) the white dwarf accretes matter from a non-degenerate companion  and explodes when it reaches the critical mass; the accreted matter can be either hydrogen or helium \cite{whel73,nomo82,han04}. ii) in the double degenerate scenario (DD) two white dwarfs merge as a consequence of the momentum losses caused by the emission of gravitational waves; the evolution of the merger is not completely understood at present and consequently it is not known at which moment the explosion will occur \cite{webb84,iben84}. iii) In the sub-Chandrasekhar scenario (SCH) it is assumed  that a C/O white dwarf, with a mass not necessarily near the critical one, accretes helium and detonates as a consequence of the shock wave generated by the ignition of the bottom of the freshly accreted layer \cite{woos94,livn95}; this helium can be directly accreted from a non-degenerate He-star or He-white dwarf, or it can accumulate in the outer layers as the product of the burning of the hydrogen that is being accreted. iv) In the white dwarf-white dwarf collision scenario (WD-WD) it is assumed that two white dwarfs collide and immediately ignite 
\cite{pakm12,kush13}. v) In the core degenerate scenario (CD) the white dwarf merges with the core of an AGB star; this case corresponds to the prompt merger in the DD scenario, and the explosion can occur at any time after the merger \cite{livi03, kash11}.

Detailed and extensive studies have already been carried out by different groups 
about the expected emission of the different supernova scenarios. This led to 
the obtention of a series of theoretical spectra for different models and 
epochs of the explosion, and to the identification of the main signatures that could be used to 
discriminate among the different models. It is important to emphasize here that our code \cite{gome98} has passed a test of consistency with other independent codes \cite{miln04}. 

\begin{wrapfigure}{l}{0.5\textwidth}
\centering
\includegraphics[width=0.5\textwidth]{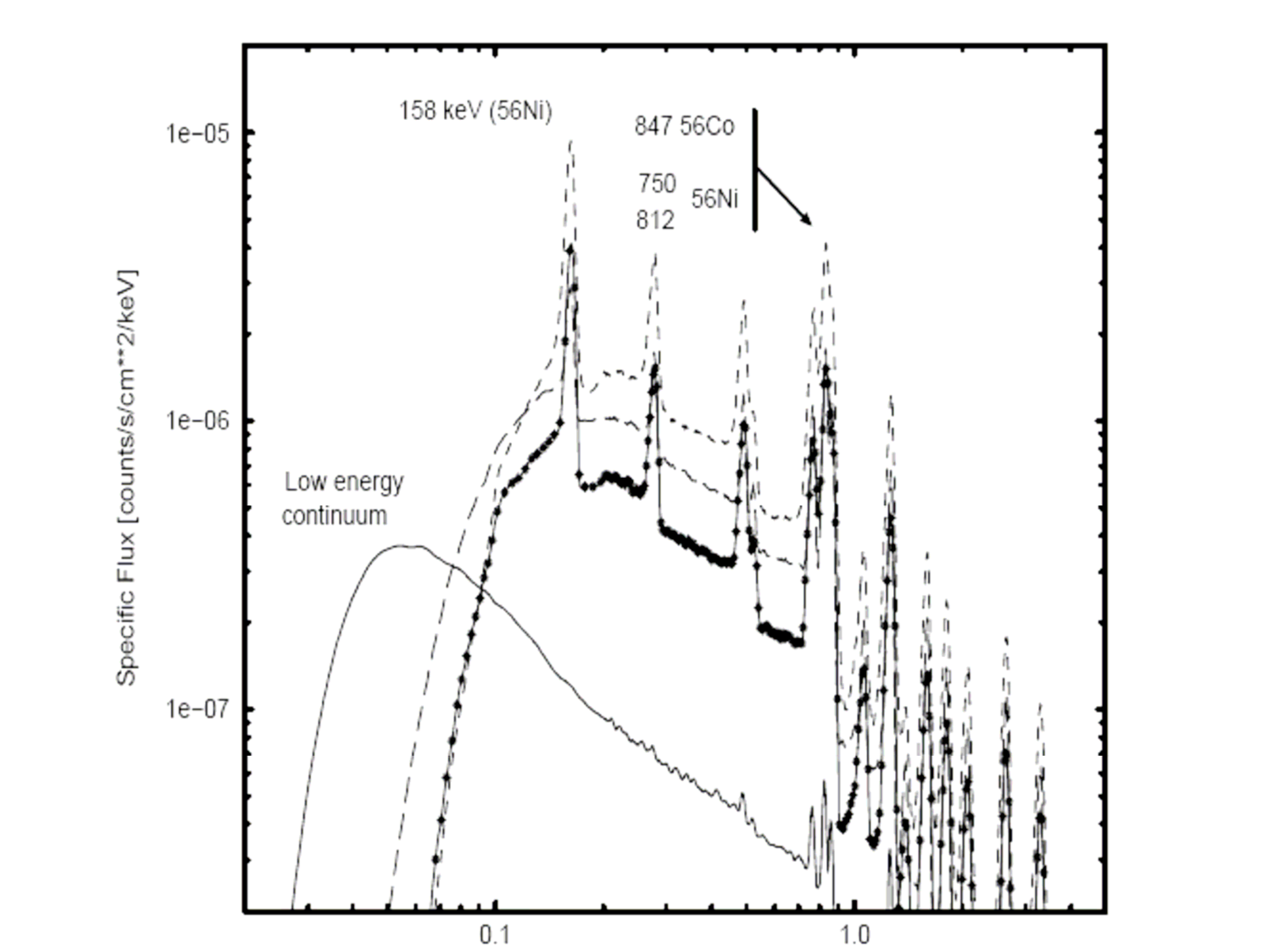}
\caption{\footnotesize{Gamma--ray spectrum for four models of SNIa explosion at 5 Mpc 20 days after the explosion. Pure deflagration model (solid line), delayed detonation model 
(long--dashed line), detonation model (dashed line) and sub--Chandrasekhar model (starred line) \cite{gome98}.}}
\label{fig1}
\end{wrapfigure}

Before and around the maximum of the visible light curve, the $\gamma$-emission is
dominated by the emission of $^{56}$Ni and  $^{56}$Co. Since the debris are still opaque, the
intensity of the emission is strongly dependent on the distribution of $^{56}$Ni within them
and changes rapidly with a time scale $t_{\rm d}$ that is a compromise between the expansion
and the decay time (T$_{1/2}\sim 6$ days) scales. For this reason the observation windows of Ni lines during this epoch have a maximum duration of $\delta t =1.26 t_{\rm d} \sim 10$ days \cite{iser13}. Because of the rapid expansion, the emergent lines are broad, typically from 3\% to 5\%, which limits the sensitivity of the instruments. During this epoch,  the spectrum (see Figure~\ref{fig1}) is dominated by the
158, 750 and 812 keV $^{56}$Ni lines as well as by the progressively growing 847 keV
 $^{56}$Co line. Because of the Doppler broadening, the 812 and the 847 keV lines blend
and their interpretation demands some care and the use of an independent
determination of the intensity of the  $^{56}$Co line at late times to disentangle both
contributions. Notice we have included te pure Chapman-Jouguet detonation model despite the fact that it does not spontaneously occur in nature. This is because the velocity of the flame only depends on the thermodynamical properties of the white dwarf and provides an useful upper bound.

As shown in figure~\ref{fig1}, twenty days after the explosion all models involving a prompt or a delayed detonation display strong lines because their high expansion rates induce a rapid decrease of the density. Lines are particularly intense for those models containing $^{56}$Ni and 
$^{56}$Co in the outer layers (pure detonation and sub--Chandrasekhar models). The maximum intensity of these lines is model dependent since it is a function of the expansion rate and of the distribution of $^{56}$Ni. Pure deflagration models only display a continuum since they efficiently Comptonize  high energy $\gamma$--rays. The shape of the continuum at low 
energies is limited in all models by the competing photoelectric absorption, which imposes a cut--off below 40--100 keV. The energy of the cut--off is determined by the chemical composition of the external layers where most of the emergent continuum is formed at this epoch. Consequently, the continuum of those models containing low Z elements in the outer layers will extend to lower energies than that of those containing high Z elements. 
Therefore, it is possible to use these differences to discriminate among the different burning modes \cite{gome98}.   

SN2014J was discovered by \cite{foss14} on January 21st 2014 in M82 ($d=3.5\pm 0.3$~Mpc). Three observation runs with \emph{INTEGRAL} were performed. During the first one, that started 16.5 days after the explosion (a.e., from now) and finished 35.2 days a.e, this emission was observed by the INTEGRAL instruments SPI \cite{dieh14,chur15,iser16} and IBIS \cite{chur15,iser16}. Although its interpretation is controversial due to the weakness of the signal and the presence of instrumental lines very near to the 158 and 812 keV lines in the SPI spectrograph \cite{iser16}, its detection is robust. The 158 keV line is critical for such purposes since the 812 keV mixes with the 847 keV line and the 750 keV one is much weaker (although this line remains fully useful for diagnostic purposes in the case of bright sources). 

\begin{figure}[h]
\begin{minipage}{17pc}
\includegraphics[width=17pc]{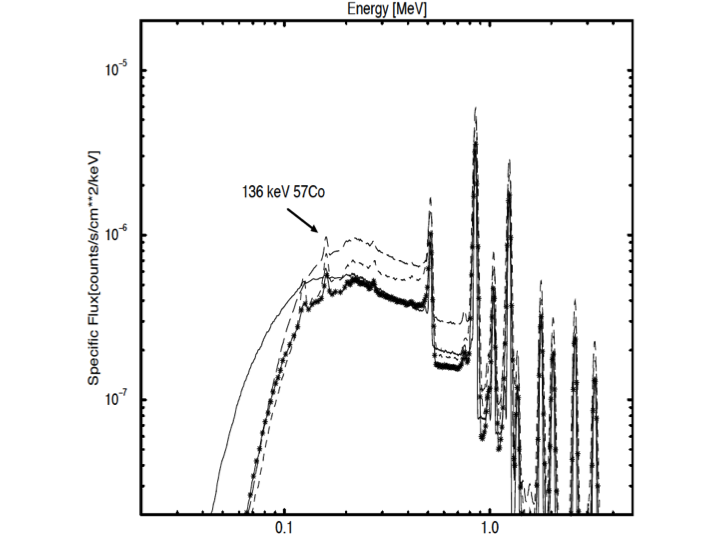}
\caption{\label{fig2} \footnotesize   Gamma--ray spectrum for four models of SNIa explosion at 5 Mpc 60 days after the explosion. Pure deflagration model (solid line), delayed detonation model 
(long--dashed line), detonation model (dashed line) and sub--Chandrasekhar model (starred line) \cite{gome98}.}
\end{minipage}\hspace{2pc}
\begin{minipage}{17pc}
\includegraphics[width=17pc]{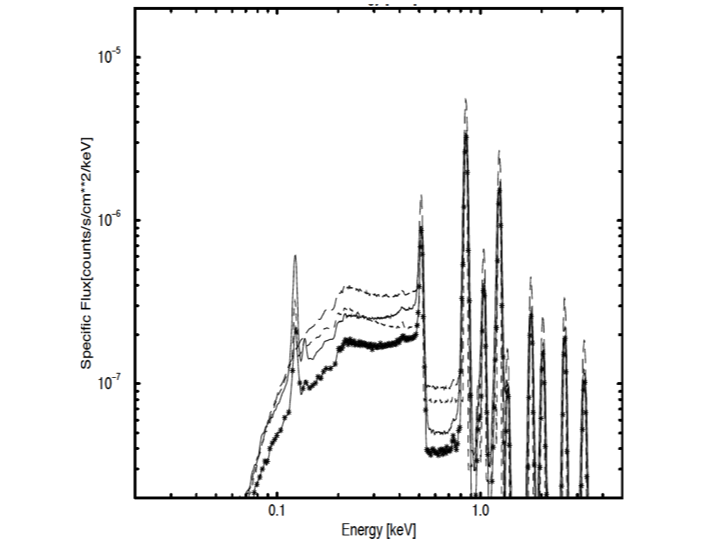}
\caption{\label{fig3} \footnotesize  Gamma--ray spectrum for four models of SNIa explosion at 5 Mpc 120 days after the explosion. Pure deflagration model (solid line), delayed detonation model (long--dashed line), detonation model (dashed line) and sub--Chandrasekhar model (starred line) \cite{gome98}.}
\end{minipage} 
\end{figure}

Two months after the explosion (see Fig.~\ref{fig2}), the $^{56}$Ni isotopes have disappeared in all the models and the emission is dominated by the $^{56}$Co lines. The 122 and 136 keV lines of $^{57}$Co are already visible although faint. At this moment, the line intensities  in the pure detonation, delayed detonation and subCH models are mainly determined by the total mass of radioactive isotopes, while the effects of the expansion rate becomes secondary. The cut--off energies of these models converge to a value of $\sim 70$ keV although it is still smaller in the deflagration model.

Four months after the explosion the ejecta are optically thin in all models (Fig.~\ref{fig3}). The continuum is faint and is dominated by the positronium annihilation component plus a contribution of photons scattered once. This contribution steeply decreases below 170 keV (the energy of a backscattered 511 keV) and a step appears at this energy. During this phase, the cut--off is associated with the characteristic spectrum of photons emitted by positronium annihilations, which is model independent. Line intensities are now proportional to the mass of the respective parent isotopes, except for the deflagration model, while the effect of the differences in expansion velocities is secondary and the position of the low energy cut--off tends to converge towards $\sim$~70 keV. At this epoch line profiles reveal the distribution in velocity of their parent isotopes in all layers of the ejecta. 

\begin{wrapfigure}{l}{0.5\textwidth}
\centering
\includegraphics[width=0.5\textwidth,clip=true, trim= 0cm 3cm 0cm 3cm]{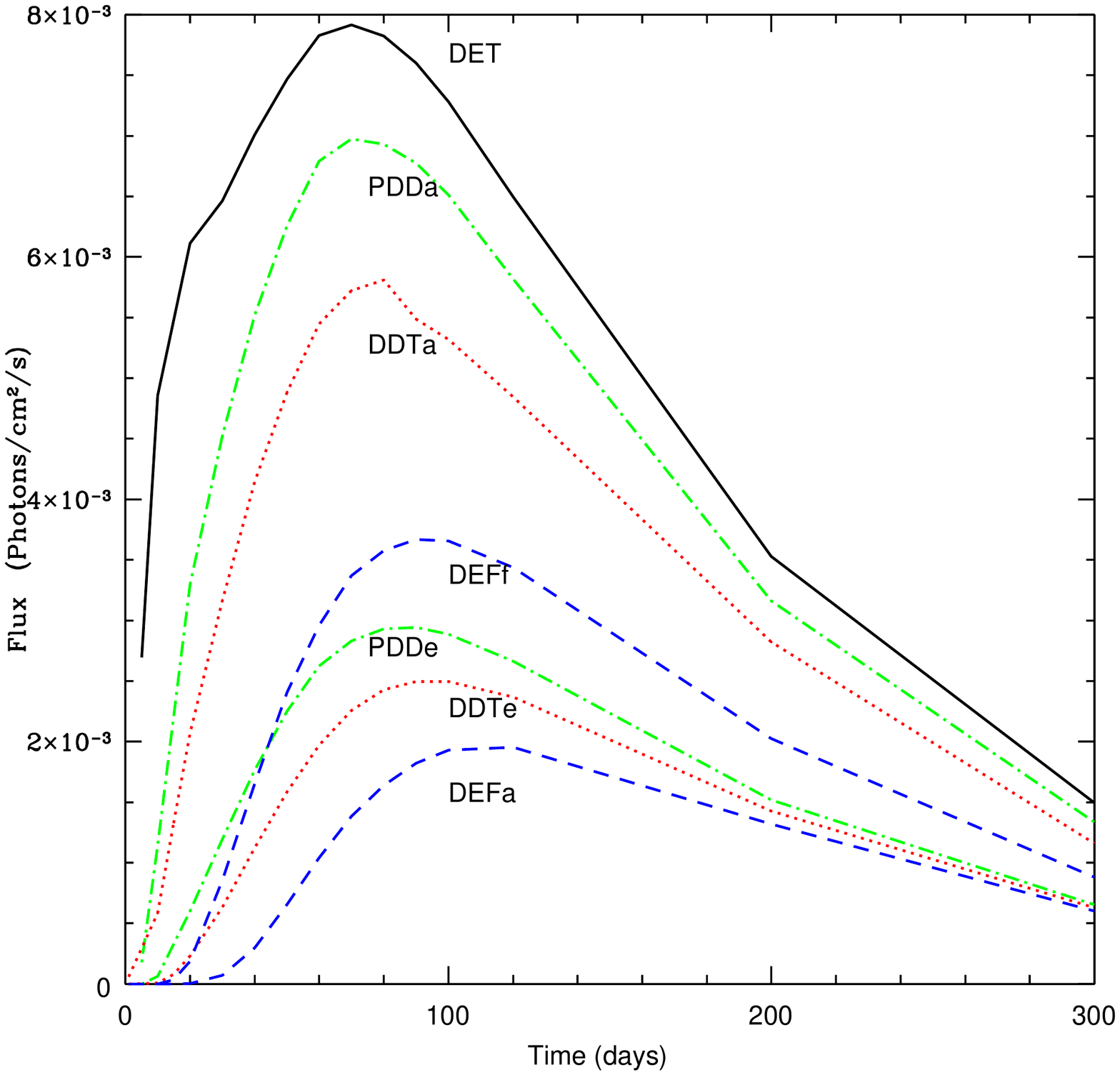}
\caption{\footnotesize{Evolution of the intensity of the 847 keV $^{56}$Co line for different burning modes. The mass of the exploding star is 1.35 M$_\odot$ and the distance is assumed to be 1 Mpc.}}
\label{fig4}
\end{wrapfigure}

As mentioned before, during  the late time epoch, when the debris are transparent to gammas, 
the emission is dominated by the 847 and 1238 keV $^{56}$Co lines, which can be used for
diagnostic purposes, as recently proved in the case of SN2014J \cite{chur14}. The measurement
of the intensity of these lines provides a direct and precise determination of the $^{56}$Co
mass, which is the main parameter that controls the Phillips relationship, the one
that allows the use of SNIa as standard candles. These two lines reach their maximum intensity
two to three months after the explosion. Since they change with a time scale of $\sim 100$~days, the observing window is only limited by programmatic reasons (Fig.~\ref{fig4}).

In all the spherically symmetric models considered here, $^{56}$Ni is buried in the inner layers and it is necessary to wait for a substantial expansion of the debris to allow the escape of non-thermalized $\gamma$-photons. Despite having important amounts of  $^{56}$Ni in the outer layers, SCH models have a similar behavior. It is important to realize here, that the viability of these SCH models has been questioned as a consequence of the severe constraints posed by the existing optical observations on the total amount of $^{56}$Ni that can be synthesized in these outer layers.

It is important to realize that all these observations are perfectly feasible if the distances at
which the events occur are short enough. This means that if the sensitivity is poor, gamma-ray observations of SNIa have
to be considered as rare events and considered as targets of opportunity. On the contrary, if
the sensitivity is good enough, a certain number (statistically speaking) of targets can be
guaranteed and SNIa could be considered as part of a core program of a mission.

For instance, with a sensitivity of $\sim 3\times 10^{-7}$~ph cm$^{-2}$ s$^{-1}$ keV$^{-1}$ (1 Msec of integration time) it would be possible to measure the intensity of the 847 keV $^{56}$Co line with a significance better than 3 sigma to a distance of $\sim 20$ and $\sim 10$ Mpc for the brightest and dimmer \emph{normal-Branch} supernovae respectively, which means that this quantity could be measured at least in $\sim 10$~SNIa in five years. This measurement could permit the calibration of the Arnett's luminosity rule ($ L_{\rm opt} \sim ~M_{\rm Ni}$) as well as that of the synthetic optical models, and would provide key data to
understand the Phillips relationship. Of course the significance and the size of this sample can
be improved just increasing the integration time. Furthermore, in close enough events it
would be possible to obtain additional information by measuring the light curve and the intensity of the lines \cite{gome98}. Since the maximum of the cobalt lines happens more than one month after the explosion, this kind of observations could be easily scheduled with the instruments on board of missions like e-ASTROGAM \cite{dean17}.

\begin{figure}[h]
\begin{minipage}{17pc}
\includegraphics[width=17pc]{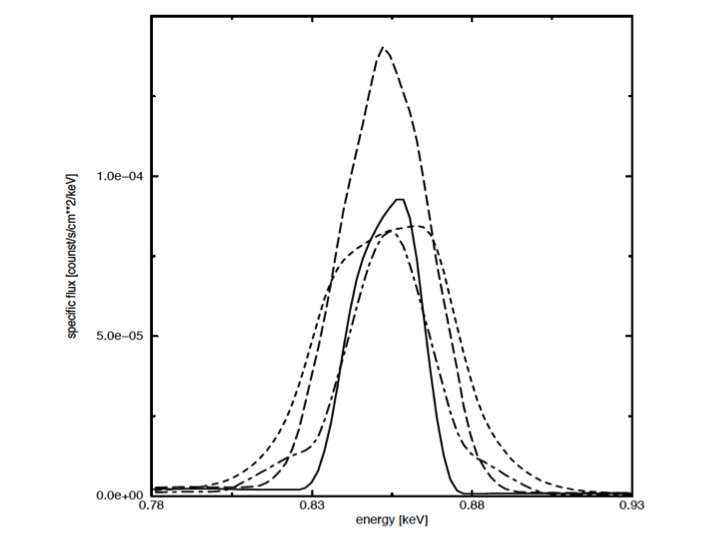}
\caption{\label{fig5} \footnotesize {Profiles of the 847 keV lines 120 days a.e. at a distance of 1 Mpc. Solid lines correspond to a deflagration model, long-dashed lines to a delayed-detonation model, dashed lines to a pure detonation model, and dot-dashed lines to a subChandrasekhar model\cite{gome98}.}}
\end{minipage}\hspace{2pc}
\begin{minipage}{17pc}
\includegraphics[width=17pc]{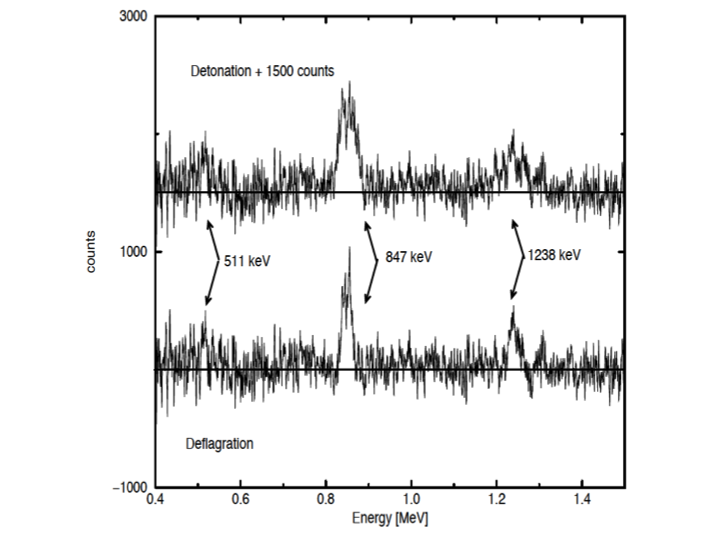}
\caption{\label{fig6} \footnotesize Simulated observational spectra for a detonation
and a deflagration SNIa at 5 Mpc (integration time= $10^6$~s). The
detonation spectrum is shifted a factor +1500 \cite{gome98}.}
\end{minipage} 
\end{figure}

\begin{wrapfigure}{l}{0.5\textwidth}
\centering
\includegraphics[width=0.5\textwidth]{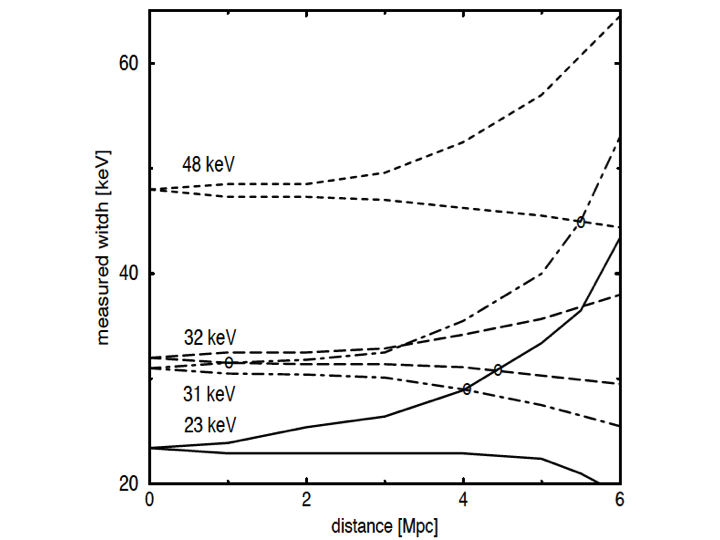}
\caption{\footnotesize{ Pairs of curves containing 90\% 
of observational line widths as a function of distance for the 847 keV line 120 days after the explosion as a consequence of the background \cite{gome98}. The line code is the same as in Fig.~\ref{fig5}}}
\label{fig7}
\end{wrapfigure}

The resolution of the line profile is another question. In principle, instruments with a spectral resolution similar to that of SPI ($\sim$ 0.2\%) should allow to identify many details of the line profiles of SNIa (Figure~\ref{fig5}) and to perform model-specific fits to the observations. Unfortunatelly, if the signal to noise is not high enough, the fluctuations of the background will hide the secondary features of the lines, (see Figure~\ref{fig6}) and in the majority of cases it will only be possible to fit the lines with gaussians. Although the lines are not exactly gaussians, the difference between the FWHM of their theoretical profiles and that of the corresponding gaussian
fit is, in all cases, below 3\%. These errors are negligible compared with observational uncertainties and hence gaussian fitting is a good technique to measure observational line
widths \cite{gome98}.

 Gomez-Gomar et al (1998) performed some simulations to evaluate the diagnostic possibilities offered by the line profile of the 847 keV line, 120 days a.e. using the expected response of SPI before launch.  
The results obtained are summarized in Figure~\ref{fig7} where for
each model a pair of curves are displayed. At every explosion
distance the pair of curves defines an interval of possible measured
widths which contains the values that would be obtained
by 90\% of observers measuring the same line at the same
distance (90\% dispersion bar). In the figure it can be appreciated
that the dispersion of the measures is 0 for an explosion
at distance 0 but it steeply grows with the explosion
distance, being larger for lines with low fluxes. This is particularly
important for the subChandrasekhar and deflagration models since they
have the lowest luminosities. For all models the distribution
of hypothetical measures is skewed. That is, the observations
are not symmetrically spread around the original line
width but there is a tendency to measure widths larger than
the original values which are indicated in the figure. As the
possible errors become more important the significance of a
measure decreases. Hence, it is necessary to adopt a quantitative
criterium which establishes the maximum distance at
which a measurement of a line width has physical meaning.
We take this distance at the point at which the width of the
dispersion bar for a line equals its original width. Assuming
this definition the distances are in the range  $\sim 5.5 - 8$~Mpc (certainly, the present response of SPI provides poorer distances). 

The detection of the early emission is more challenging but much more rewarding since it can
provide direct information about the development of the explosion and could confirm or not
the presence of $^{56}$Ni in the outer layers as well as to determine if this presence is a general
property of normal supernovae or just an anomaly. For instance, the analysis of the 158 KeV line by both SPI and ISGRI on board of INTEGRAL strongly suggests the presence of $^{56}$Ni in the
outer layers of SN2014J \cite{iser16}, an otherwise normal SNIa, and the absence of this
isotope in the optical spectrum during this epoch has been considered as one of the most
characteristic properties of SNIa! If the presence of $^{56}$Ni in the outer layers were confirmed it would be possible to consider scenarios involving the accretion of helium and its off-center ignition, as well as to obtain insight on the propagation of
flames in the outer envelope of the exploding star. Furthermore, the inclusion of an additional
source of non-thermalized gamma photons around 1 MeV could have a strong influence on
the properties of the cosmic MeV background.  The impact of this new circumstance on using these events as cosmological tools remains to be evaluated. In any case, the ability of measuring the 158 keV line is critical. 

\section{Conclusions}
The observation of SN2014J in M82 with $\gamma$--rays has been  a difficult 
task despite the short distance, $\sim 3.5$~Mpc, at which it exploded. It is obvious that a noticeable improvement of the sensitivity of detectors in the region of the MeV is absolutely necessary. Missions like eASTROGAM \cite{dean17} or detectors like the Laue lens \cite{vonb12} are absolutely necessary to explore this energy region. The information that these instruments  could provide is of fundamental importance not only to understand thermonuclear supernovae, but also other events like novae or the MeV background.

\section*{Acknowledgements}

This  work was  supported by the  MINECO-FEDER grants ESP2013-47637-P \& ESP2015-66134-R (JI), AYA2015-63588-P (EB), by the grant 2009SGR315 and the CERCA program of the Generalitat de Catalunya (JI).

\end{document}